\documentclass[usegraphicx]{emulateapj} 
\usepackage{longtable} 
\usepackage{lscape} 
\usepackage{natbib} 

\newcommand{\placefigone}{ 
\begin{figure}[ht]  
\begin{center} 

\plotone{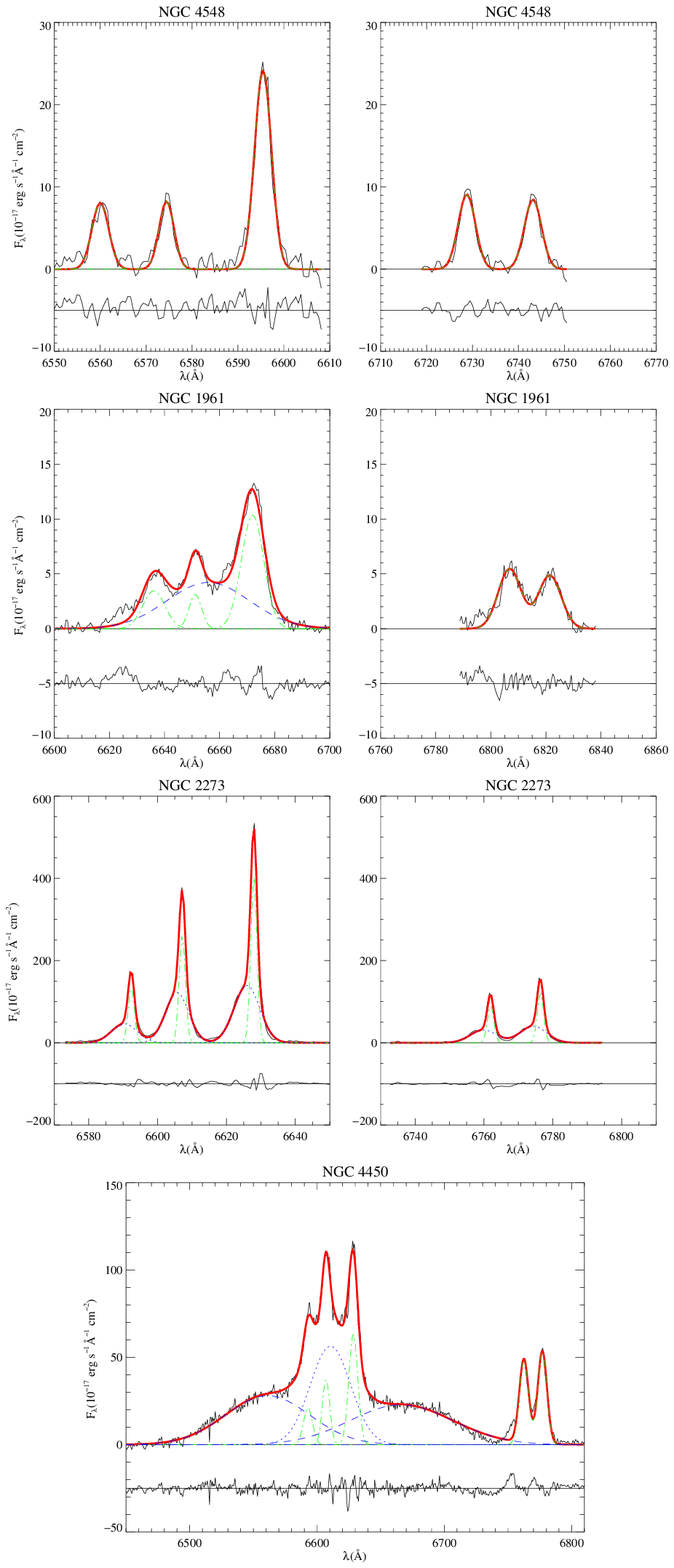}
\end{center} 
 \caption{Few examples of continuum-subtracted central G750M spectra
   from our spectral atlas illustrating the various fitting strategies
   adopted to match \ha, \niipg\ and \siipg\ emission lines. In each
   panel the red line show the overall line blend, whereas the green
   dashed-dotted lines and blue dotted or dashed lines show the
   adopted narrow and broad Gaussian components, respectively. Shown
   are also the fit residuals, offset for better visibility. For
   NGC~4548 the nebular emission could be match with single Gaussian
   profiles, for NGC~1961 we needed to add a broad \ha\ component,
   whereas for NGC~2273 an additional broad and blue-shifted component
   was needed to macth the profile of all lines. For NGC~4450 we added
   two extremely broad Gaussians shoulders offset from the center by
   several thousands \kms, in addition to a more typical broad \ha\
   component \citep[see also][for a match to the double-peaked profile
   of this LINER 1.9 nucleus]{Ho2000}.}
\label{fig:fitexample}  
\end{figure}  
}

\newcommand{\placefigtwo}{  
\begin{figure}[ht]   
\begin{center}  
\plotone{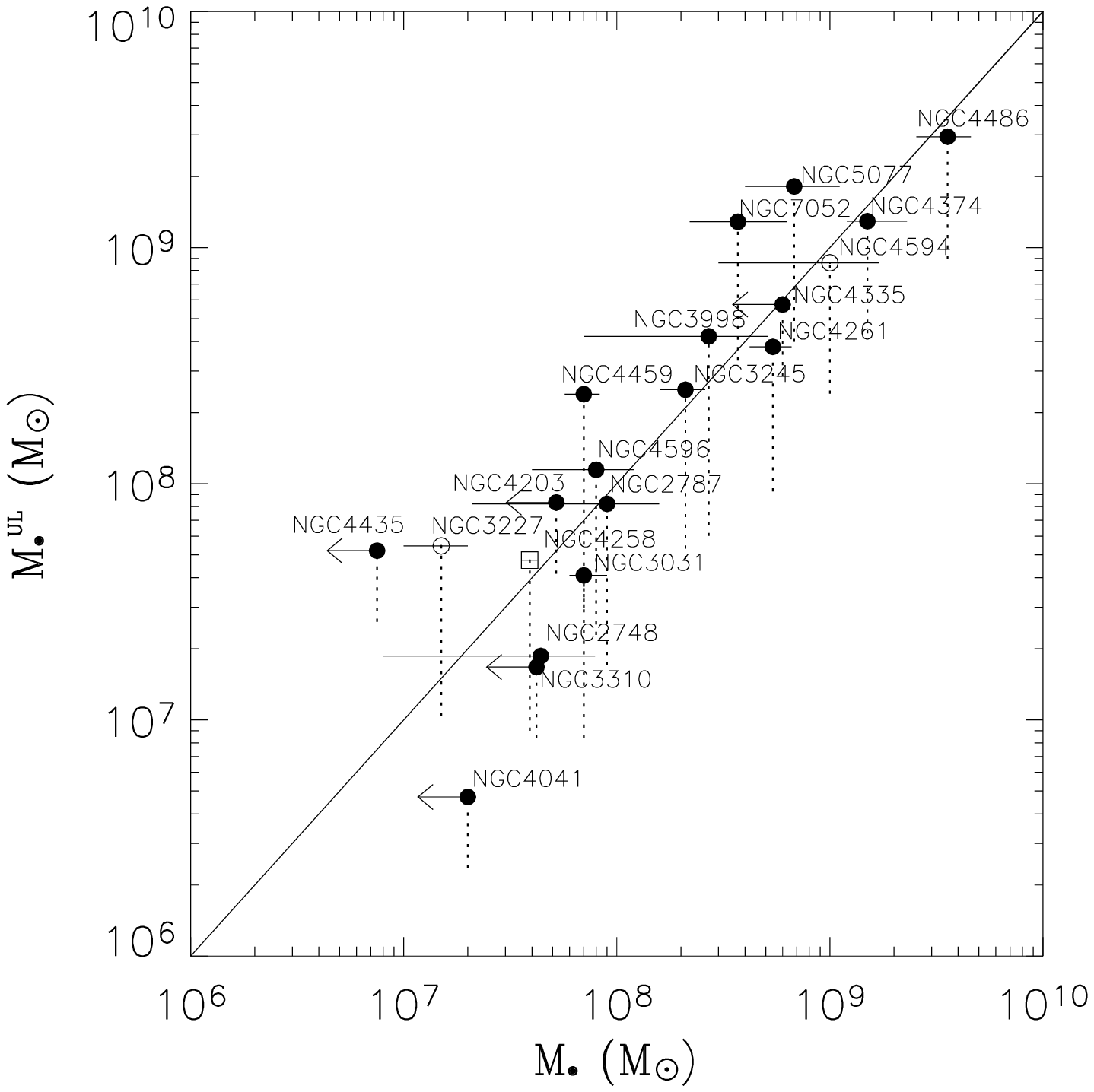} 
\end{center} 
  \caption{Comparison between our \mbh\ upper limits and accurate
    measurements of \mbh\ based on the resolved kinematics of gas
    (filled circles), stars (open circles), and water masers (open
    square) available in the literature. Leftward arrows indicate an
    upper constrain rather than a definite value for \mbh. The upper
    and lower edges of the dotted lines correspond to the \mbh\ values
    that we estimated assuming an inclination of $i=33^\circ$ and
    $81^\circ$ for the unresolved Keplerian disk, respectively.}
\label{fig:ULcomparison}   
\end{figure}   
}

\newcommand{\placefigthree}{  
\begin{figure}[ht]   
\begin{center}  
\epsscale{.9}
\plotone{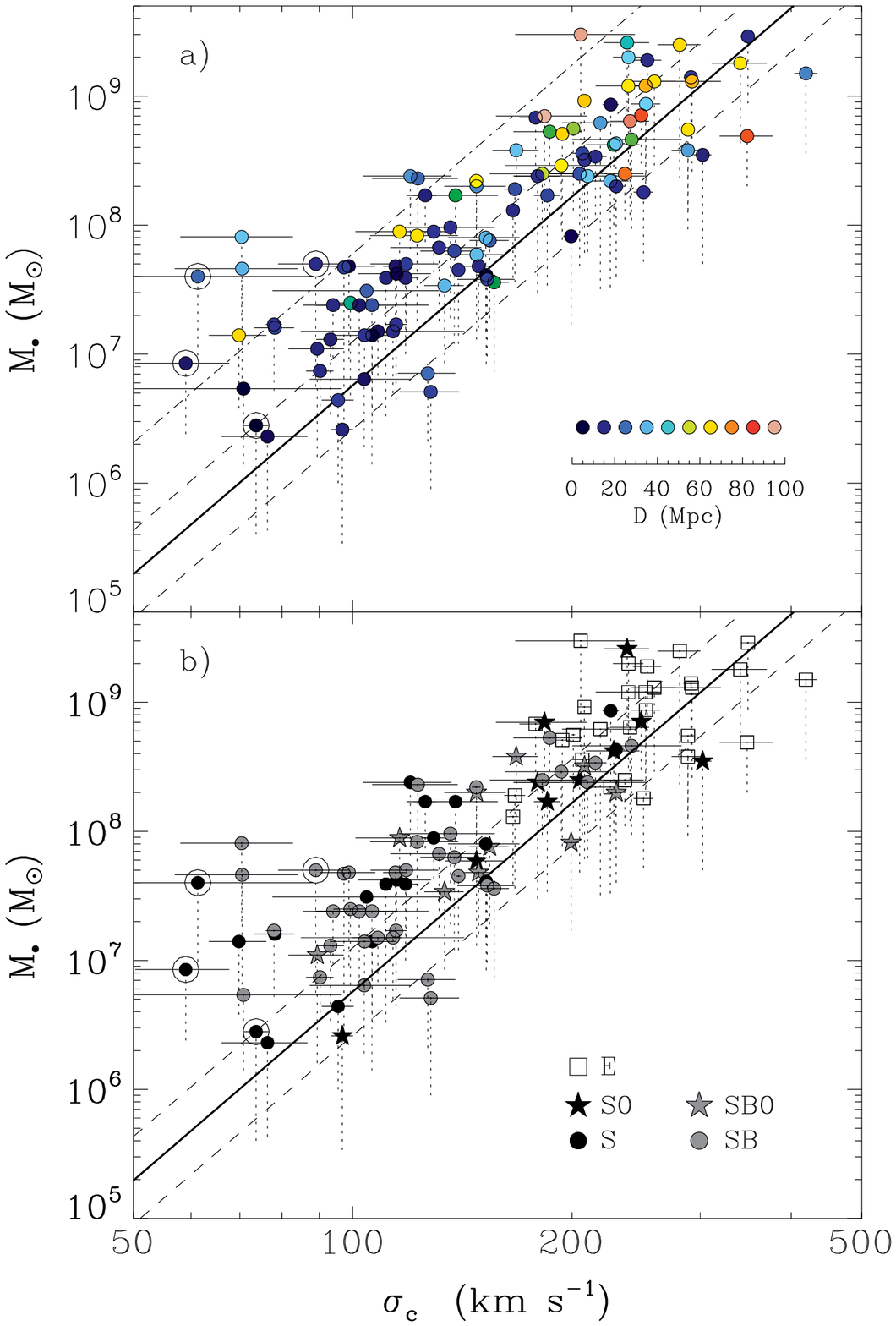} 
\end{center}  
  \caption{Comparison between our \mbh\ upper limits and \msc\
    relation by \cite{Ferrarese2005} (thick line) as a function of
    galaxy distance {\em (a)} and morphological type {\em (b)}. The
    upper and lower edges of the dotted lines correspond to \mbh\
    values estimated assuming an inclination of $i=33^\circ$ and
    $81^\circ$ for the unresolved Keplerian disk, respectively. Large
    circles mark galaxies with $\sigma_{\rm c} < 90$ \kms\ that host a
    nuclear star cluster. The dashed lines show to the 1$\sigma$ (0.34
    dex) scatter in \mbh. Additionally, to follow the discussion in
    \S\ref{subsec:coeff} and \S\ref{subsec:low}, the dot-dashed line
    shows the 3$\sigma$ (1.02 dex) scatter above the \ms\ relation
    whereas the open circles point to objects where a nuclear cluster
    is present.}
\label{fig:ULmsigmaFF05}  
\end{figure}  
}  
  
\newcommand{\placefigfour}{  
\begin{figure}[ht]   
\begin{center}
\epsscale{.9}
\plotone{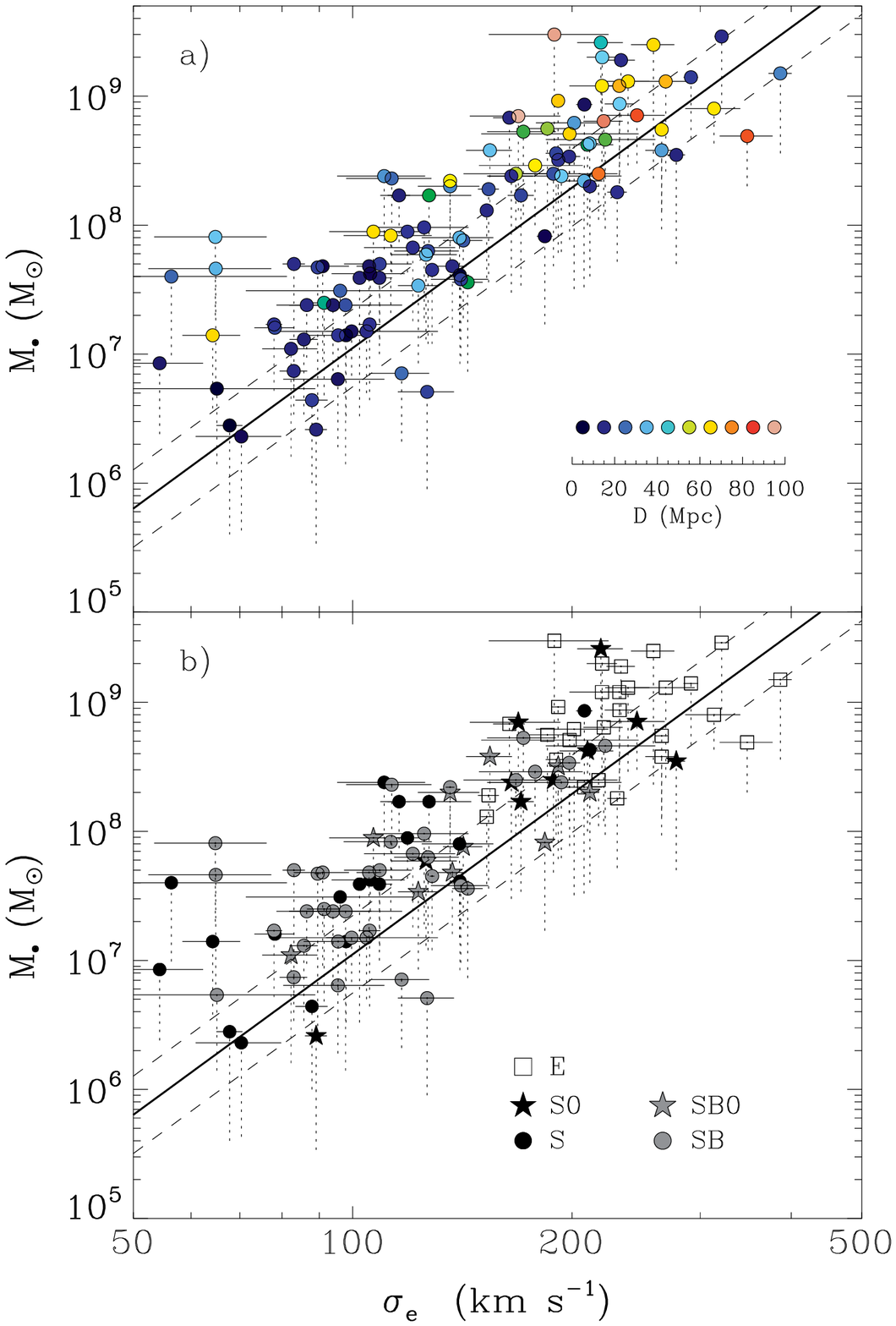} 
\end{center}  
 \caption{Same as Fig. \ref{fig:ULmsigmaFF05} but now showing the
    comparison between our \mbh\ upper limits and the \mse\ relation
    of \citet{Lauer2007b}}
\label{fig:ULmsigmaLauer07}  
\end{figure}  
}  

\newcommand{\placefigfive}{  
\begin{figure}[ht]   
\begin{center}
\epsscale{.95}
\plotone{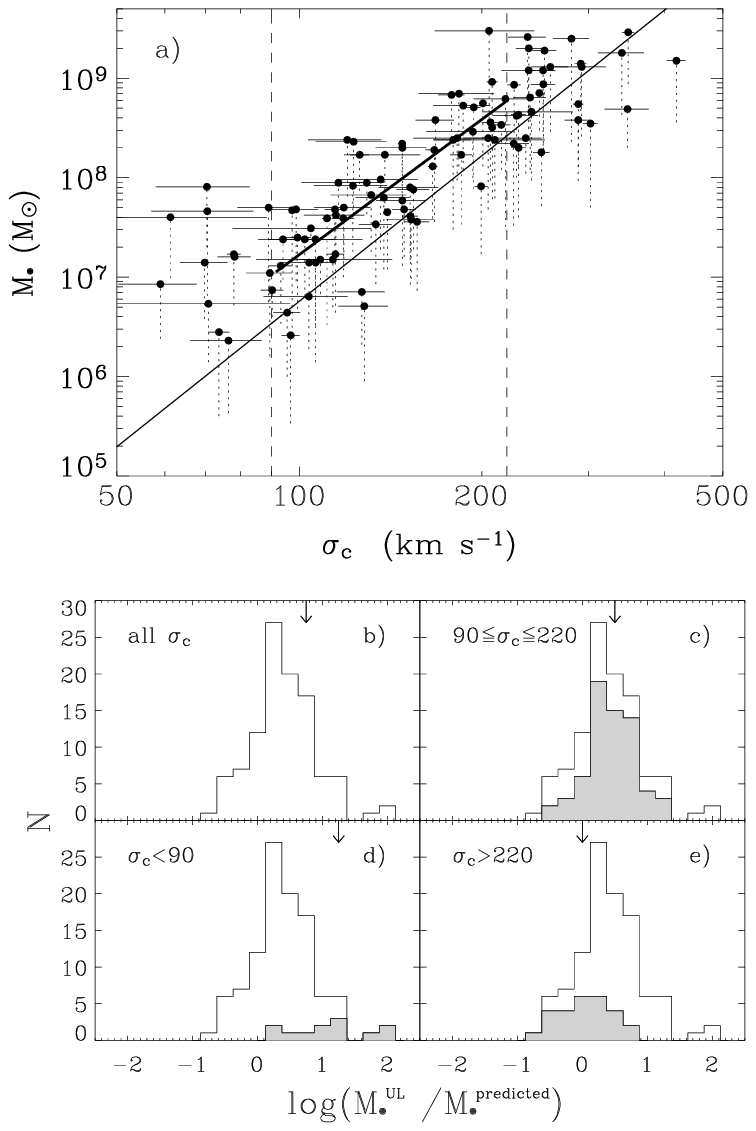} 
\end{center}  
 \caption{{\em Upper panel (a):} Comparison between the linear fit to
   the \mbh\ upper limits in the range $90 \leq \sigma_{\rm c} \leq
   220$ \kms\ (thick line) and the \msc\ relation by \citet[][thin
   line]{Ferrarese2005}. Our linear fit to the \mbh\ upper limits in
   the $90 \leq \sigma_{\rm c} \leq 220$ \kms\ range (shown by the
   vertical dashed lines) delivers a best-fitting slope of
   $4.52\pm0.41$. {\em Lower panels:} Distribution of the ratios
   between the measured upper limits and the values of \mbh\ expected
   from the \msc\ relation by \cite{Ferrarese2005} for {\em (b)} all
   the sample galaxies, {\em (c)} the galaxies with $90 \leq
   \sigma_{\rm c} \leq 220$ \kms , {\em (d)} with $\sigma_{\rm c} <
   90$ \kms , and {\em (e)} with $\sigma_{\rm c} > 220$ \kms . The
   median of each distribution is marked by an arrow.}
\label{fig:histograms}  
\end{figure}  
}

\newcommand{\placefigsix}{  
\begin{figure}[ht]   
\begin{center}
\epsscale{1.2}
\plotone{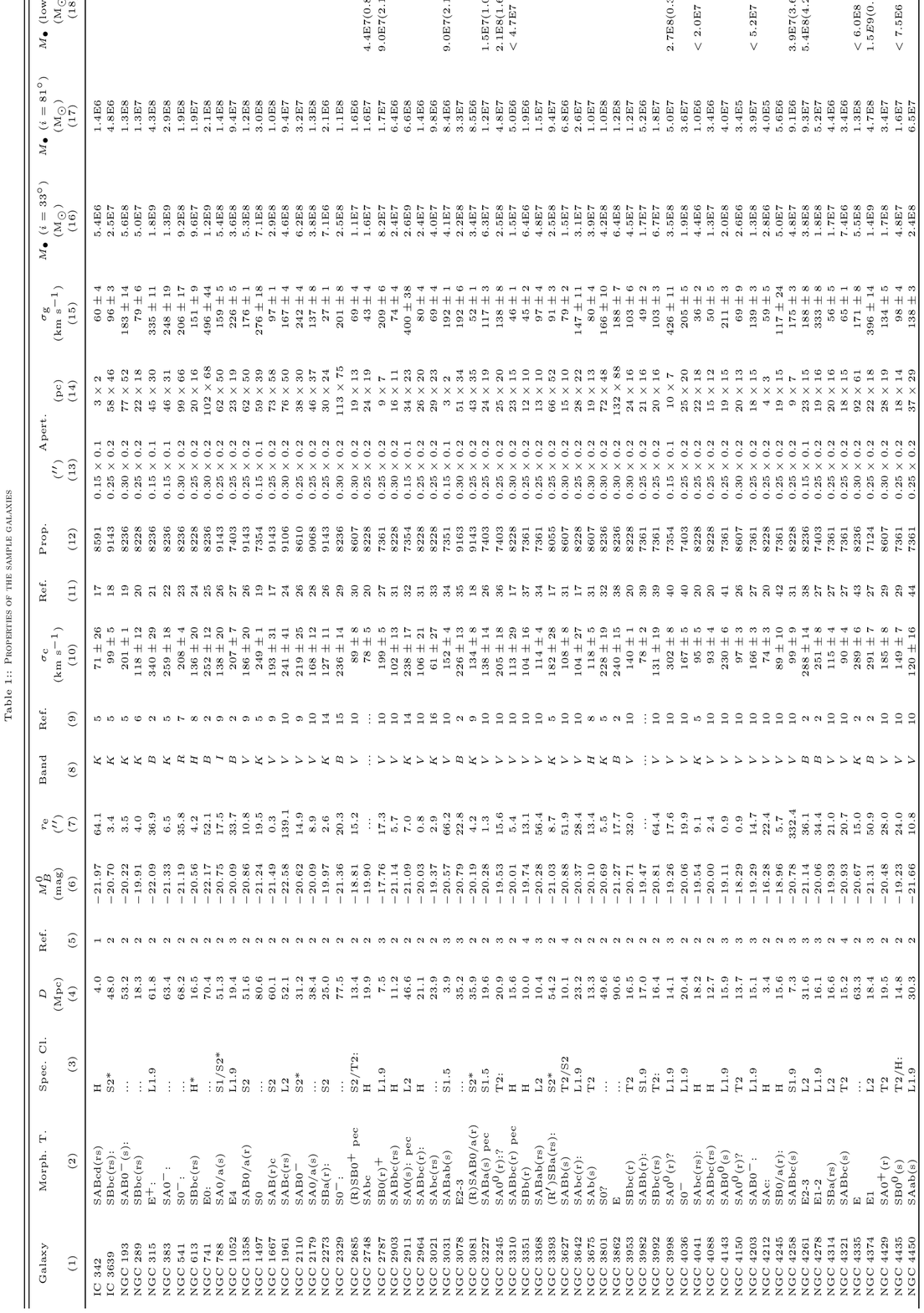} 
\end{center}  
\label{tab:allMBH}
\end{figure}  
}

\newcommand{\placefigseven}{  
\begin{figure}[ht]   
\begin{center}
\epsscale{1.2}
\plotone{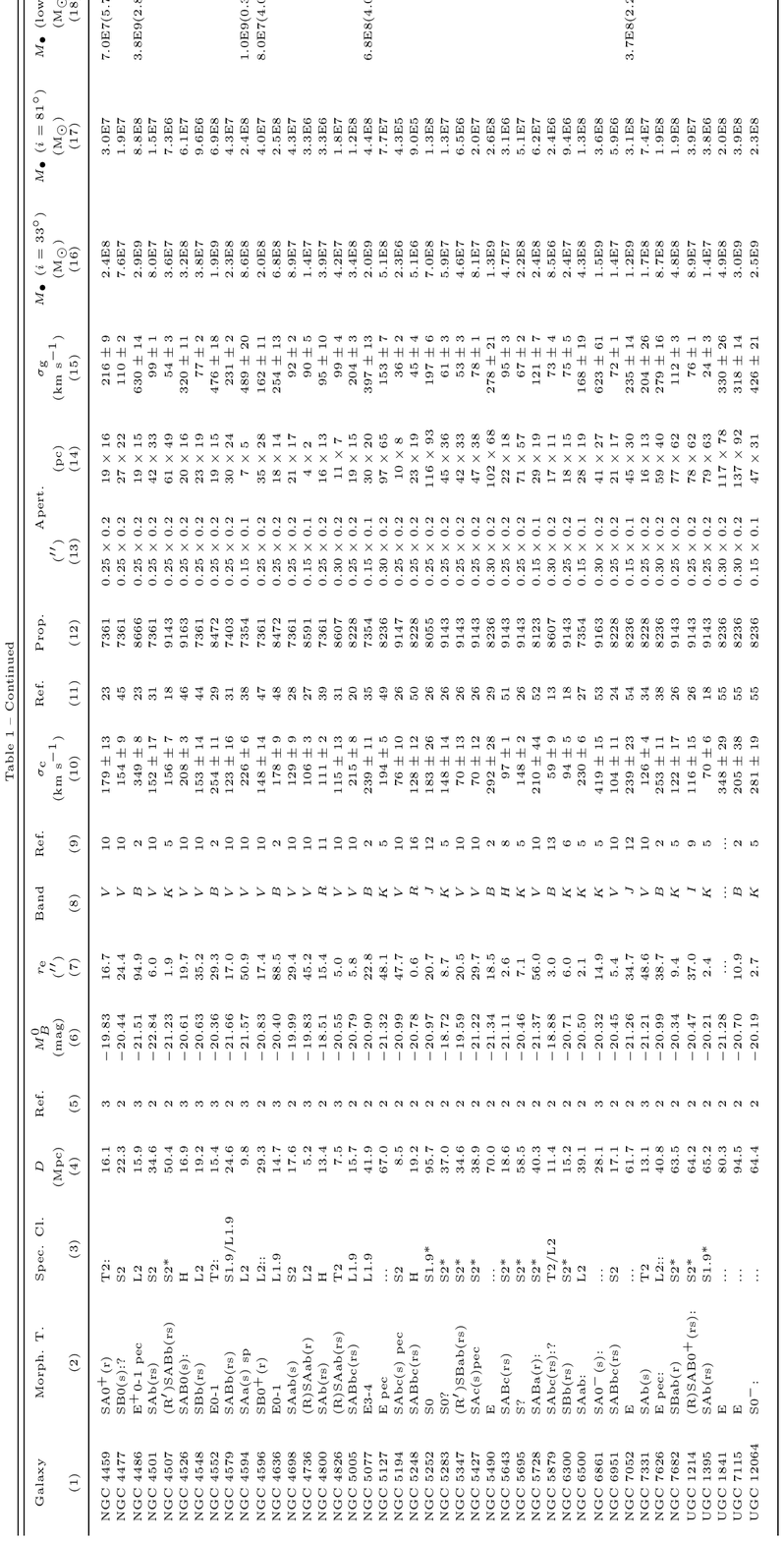} 
\end{center}  
\label{tab:allMBH}
\end{figure}  
}

\def\kms{$\mathrm{km\;s}^{-1}$}  
\def\kmsmpc{$\mathrm{km\;s^{-1}\;Mpc^{-1}}$}  
  
\def\sigmac{$\sigma_{\rm c}$} 
\def\sigmae{$\sigma_{\rm e}$} 
  
\def\msun{M$_{\odot}$}  
  
\def\mbh{$M_\bullet$}  
  
\def\ms{$M_\bullet-\sigma$}  
\def\msc{$M_\bullet-\sigma_{\rm c}$}  
\def\mse{$M_\bullet-\sigma_{\rm e}$}  

\def\nii   {[{\sc N$\,$ii}]}  
  
\def\niig  {[{\sc N$\,$ii}]$\,\lambda6583$}  
\def\niipg {[{\sc N$\,$ii}]$\,\lambda\lambda6548,6583$}  
\def\ha{H$\alpha$}  
\def\sii   {[{\sc S$\,$ii}]}

\def\siipg {[{\sc S$\,$ii}]$\,\lambda\lambda6716,6731$}

\defcitealias{Andredakis1994}{11} 
\defcitealias{Atkinson2005}{57} 
\defcitealias{Baggett1998}{10} 
\defcitealias{Balcells1995}{55} 
\defcitealias{Barth2001}{61}   
\defcitealias{Barth2002}{27} 
\defcitealias{Batcheldor2005}{20} 
\defcitealias{Bender1994}{48}  
\defcitealias{Bernardi2002}{23} 
\defcitealias{Bettoni1997}{47}  
\defcitealias{Bonfanti1995}{25}  
\defcitealias{Bower1998}{66}  
\defcitealias{Carollo1993}{35} 
\defcitealias{Coccato2006}{68}  
\defcitealias{Corsini1999}{28}  
\defcitealias{Davies1987}{21} 
\defcitealias{Davies2006}{60}  
\defcitealias{deFrancesco2006}{63} 
\defcitealias{deFrancesco2008}{71} 
\defcitealias{deSouza2004}{7} 
\defcitealias{Devereux2003}{59} 
\defcitealias{diNella1995}{32} 
\defcitealias{Dumas2007}{50}  
\defcitealias{FalconBarroso2002}{13}  
\defcitealias{FalconBarroso2006}{42} 
\defcitealias{Ferrarese1996}{67}  
\defcitealias{Fisher1995}{15}  
\defcitealias{Fisher1997}{40}  
\defcitealias{Freedman2001}{4} 	     
\defcitealias{GarciaRissmann2005}{18} 
\defcitealias{Gu2006}{51} 
\defcitealias{Heraudeau1998}{31} 
\defcitealias{Heraudeau1999}{33} 
\defcitealias{Jarvis1988}{45} 
\defcitealias{Kassin2006}{6} 
\defcitealias{Kormendy1988}{70} 
\defcitealias{Koprolin2000}{53}
\defcitealias{Laurikainen2004}{8} 
\defcitealias{Laurikainen2005}{14} 
\defcitealias{Macchetto1997}{69}  
\defcitealias{Marconi2003a}{12}  
\defcitealias{Marconi2003b}{64}  
\defcitealias{Miyoshi1995}{65}  
\defcitealias{Nelson1995}{26} 
\defcitealias{NoelStorr2007}{49} 
\defcitealias{Pastorini2007}{62} 
\defcitealias{Proctor2002}{46} 
\defcitealias{RC3}{2}  
\defcitealias{Sarzi2001}{58}  
\defcitealias{Sarzi2002}{39}  
\defcitealias{Scarlata2004}{16} 
\defcitealias{Schechter1983}{24} 
\defcitealias{Simien1997a}{22} 
\defcitealias{Simien1997b}{29} 
\defcitealias{Simien1997c}{30} 
\defcitealias{Simien1998}{36} 
\defcitealias{Simien2002}{41} 
\defcitealias{Smith2000}{38} 
\defcitealias{Terlevich1990}{17} 
\defcitealias{Tonry1981}{55} 
\defcitealias{Tonry2001}{3} 
\defcitealias{Tully1988}{1} 
\defcitealias{vandenBosch1995}{54} 
\defcitealias{vanderMarel1998}{72} 
\defcitealias{VegaBeltran2001}{34} 
\defcitealias{VerdoesKleijn2002}{43} 
\defcitealias{Wagner1988}{52} 
\defcitealias{Wegner2003}{19} 
\defcitealias{Whitmore1979}{37} 
\defcitealias{Whitmore1981}{44} 
\defcitealias{Xanthopoulos1996}{9}  
\defcitealias{2MASS}{5}

\shorttitle{Upper limits on the masses of 105 Black Holes} 
\shortauthors{Beifiori et al.}

\begin{document}  
  
\topmargin = 1cm
 
\title{Upper limits on the masses of 105 supermassive black holes from
  HST/STIS archival data \footnotemark[1]}

\footnotetext[1]{Based on observations with the
  NASA/ESA {\em Hubble Space Telescope} obtained at STScI, which is
  operated by the Association of Universities for Research in
  Astronomy, Incorporated, under NASA contract NAS5-26555.}

\author{A. Beifiori\altaffilmark{2},  
        M. Sarzi\altaffilmark{3},  
        E.~M. Corsini\altaffilmark{2},  
        E. Dalla Bont\`a\altaffilmark{2},  
        A. Pizzella\altaffilmark{2},   
        L. Coccato\altaffilmark{4}, and  
        F. Bertola\altaffilmark{2}}

\altaffiltext{2}{Dipartimento di Astronomia, Universit\`a di Padova, 
  vicolo dell'Osservatorio 3, I-35122 Padova, Italy; 
  alessandra.beifiori@unipd.it; enricomaria.corsini@unipd.it;
  elena.dalabonta@unipd.it; alessandro.pizzella@unipd.it;
  francesco.bertola@unipd.it} 
\altaffiltext{3}{Centre for Astrophysics Research, University of 
  Hertfordshire, College Lane, Hatfield AL10 9AB, UK;
  m.sarzi@herts.ac.uk} 
\altaffiltext{4}{Max-Planck-Institut f\"{u}r extraterrestrische 
  Physik, Giessenbachstrasse 1, D-85748 Garching bei M\"{u}nchen, 
  Germany; lcoccato@mpe.mpg.de}

\begin{abstract}    
  
Based on the modeling of the central emission-line width measured over
sub-arcsecond apertures with the {\em Hubble Space Telescope}, we
present stringent upper bounds on the mass of the central supermassive
black hole, \mbh, for a sample of 105 nearby galaxies ($D<100\,\rm
Mpc$) spanning a wide range of Hubble types (E $-$ Sc) and values of
the central stellar velocity dispersion, \sigmac\ (58 $-$ 419 \kms).
For the vast majority of the objects the derived \mbh\
upper limits run parallel and above the well-known \msc\ relation
independently of the galaxy distance, suggesting that our nebular
line-width measurements trace rather well the nuclear gravitational
potential.
For values of \sigmac\ between 90 and 220 \kms\ the 68\% of our upper limits
falls immediately above the \msc\ relation without exceeding the
expected \mbh\ values by more than a factor 4.1.
No systematic trends or offsets are observed in this \sigmac\ range as
a function of the galaxy Hubble type or with respect to the presence
of a bar.
For 6 of our 12 \mbh\ upper limits with \sigmac\ $< 90$ \kms\ our
line-width measurements are more sensitive to the stellar contribution
to the gravitational potential, either due to the presence of a
nuclear stellar cluster or because of a greater distance compared to
the other galaxies at the low-\sigmac\ end of the \msc\ relation.
Conversely, our \mbh\ upper bounds appear to lie closer to the
expected \mbh\ in the most massive elliptical galaxies with values of
\sigmac\ above 220 \kms.
Such a flattening of the \msc\ relation at its high-\sigmac\ end would
appear consistent with a coevolution of supermassive black holes and
galaxies driven by dry mergers, although better and more consistent
measurements for \sigmac\ and $K$-band luminosity are needed for these
kind of objects before systematic effects can be ruled out.

\end{abstract}   
  
  
%\date{{\it Draft version on \today}}   
  
\section{Introduction}  
\label{sec:introduction}  

Supermassive black holes (SMBHs) have now been discovered in the
center of a sufficiently large number of nearby galaxies to probe
possible links between the masses of SMBHs (\mbh) and the global
properties of their host galaxies.
In fact, it has emerged that \mbh\ correlates with the luminosity
\citep{Kormendy1995, Marconi2003a}, mass \citep{Magorrian1998,
Haering2004}, stellar velocity dispersion \citep{Ferrarese2000,
Gebhardt2000, Tremaine2002, Ferrarese2005}, light concentration
\citep{Graham2001}, and gravitational binding energy \citep{Aller2007}
of the host-galaxy spheroidal component, i.e., the entire galaxy in
the case of elliptical galaxies or the bulge of disk galaxies.
In light of these findings it is now widely accepted that the
mass-accretion history of a SMBH is tightly related through feedback
to the formation and evolution of the host spheroid
\citep[e.g.,][]{Silk1998, Haehnelt2000, DiMatteo2005}, some studies
having suggested a link with the mass of the dark-matter halo
\citep{Ferrarese2002, Pizzella2005}.
  
The slope and scatter of all these correlations remain quite uncertain
\citep{Novak2006}, however, in particular since they are still based
on a limited sample of galaxies with reliable \mbh\ that is biased
towards early-type systems and that is clustered around a rather
limited range of stellar velocity dispersion ($\sigma$), approximately
between $150$ and $250$ \kms.
Given the great theoretical interest spurred by these findings, there
is a pressing need to acquire better \mbh\ statistics, both in terms
of the number of targets and in terms of broadening the range of
parent galaxies, in particular towards spiral galaxies.
  
Secure \mbh\ measurements in external galaxies are traditionally
obtained through the modeling of the stellar and/or gaseous
kinematics, most often as derived using {\em Hubble Space Telescope
(HST)} observations in the optical domain.
The advent of adaptive-optics systems working at near-infrared
wavelengths has led to more stellar-dynamical measurements of \mbh\
from the ground \citep[][]{Houghton2006, Nowak2007}. Yet, such
measurements are still quite expensive, not only because good-quality
measurements of the stellar kinematics in the near-infrared require
relatively long observations, but also because proper modeling of the
stellar kinematics in the immediate vicinity of SMBHs needs robust
constraints on the importance of radial orbits and thus additional
large-scale observations, possibly with integral-field spectroscopy
\citep{Valluri2004, Cappellari2005}.
Water-masers have provided the most accurate extragalactic \mbh\
measurements to date, but such gaseous systems are exceedingly rare
\citep{Braatz1994, Greenhill2003}.
The modeling of the nuclear ionized-gas kinematics has also led to
accurate \mbh\ measurements \citep[e.g.,][]{Barth2001,
DallaBonta2008}, and at a relatively cheap cost in terms of
observation time compared to stellar-dynamical \mbh\ determinations
\citep[e.g.,][]{Verolme2002, Gebhardt2003}. Yet, only an handful of
the objects targeted by HST turned out to have sufficiently regular
gas velocity fields for the purpose of modeling \citep[][]{Sarzi2001}.
Thus, unless a large number of galaxies pre-selected to have regular
nuclear gas kinematics \citep[for instance following][]{Ho2002} is
observed with HST if and when the Space Telescope Imaging Spectrograph
(STIS) is successfully refurbished, it is unlikely that the number of
galaxies with secure \mbh\ measurements will increase dramatically in
the near future.
  
The HST Science Archive already contains an untapped resource that can
be used to better constrain the black-hole mass budget across the
different morphological types of galaxies, which consists of the vast
number of the STIS spectra from which a central emission-line width
can be measured.
The modeling of this kind of data can indeed lead to tight upper
limits on \mbh, as first shown by \citet{Sarzi2002}.
For this reason we started a program aimed at deriving \mbh\ upper
limits based on HST spectra for the largest possible number of
galaxies and a wide range of morphological types.
In this paper we present the results based on a sample of 105 nearby
galaxies for which STIS/G750M spectra in the \ha\ region and
measurements of the stellar velocity dispersion were available from
the HST archive and in the literature, respectively.
Although we will be able only to set an upper limit on the \mbh\ of
our galaxies, the lack of exact measurements will be compensated for
by the large number of upper limits when studying SMBH mass-host
galaxy relationships.
  
The paper is organized as follows. In \S~\ref{sec:analysis} we
describe our sample selection and the measurement of central
emission-line width, before briefly describing our modeling. We will
then present our results and discuss our findings in the context of
the \ms\ relation between the SBHM mass and central stellar velocity
dispersion of the host spheroid in \S~\ref{sec:results}.
  
\section{Data compilation and analysis}  
\label{sec:analysis}  
  
\subsection{Sample selection and data reduction}  
\label{sec:sample}  
  
In order to assemble the largest possible sample of homogeneous
measurements of the central emission-line width, we queried the HST
Science Archive for objects with STIS spectra obtained with the G750M
grating through either the $0\farcs1 \times 52''$ or the $0\farcs2
\times 52''$ slit placed across the galaxy nucleus, most often at
random position angles. This is indeed by far the most common
configuration in the archive, which always gives us access to the
\niipg, \ha\ and \siipg\ emission lines.
Limiting ourselves to galaxies within 100 Mpc to minimize the impact
of the stellar potential on our results, we retrieved data for 177
galaxies spanning the whole range of morphological types.
When available, galactic distances were adopted from measurements
based either on surface-brightness fluctuations \citep{Tonry2000,
Tonry2001}, Cepheid variables \citep{Freedman2001} or from
\citet{Tully1988}. Otherwise we used the weighted mean recessional
velocity corrected to the reference frame defined by the microwave
background radiation from \citet[][RC3 hereafter]{RC3} to derive the
distance to our sample galaxies by assuming $H_0=75$ \kmsmpc,
$\Omega_{\rm m}=0.3$, and $\Omega_{\Lambda}=0.7$ . The median distance
of the sample galaxies is 21.4 Mpc.
  
The archival spectra were reduced using IRAF\footnote{IRAF is
distributed by NOAO, which is operated by AURA Inc., under contract
with the National Science Foundation} and the STIS reduction pipeline
maintained by the Space Telescope Science Institute
\citep{Dressel2007}. The basic reduction steps included overscan
subtraction, bias subtraction, dark subtraction, and flatfield
correction.
Different spectra obtained for the same slit position were aligned
using IMSHIFT and knowledge of the adopted shifts along the slit
position. Cosmic ray events and hot pixels were removed using the task
LACOS\_SPEC by \citet{vanDokkum2001}.  Residual bad pixels were
corrected by means of a linear one-dimensional interpolation using the
data quality files and stacking individual spectra with IMCOMBINE,
This allowed to increase the signal-to-noise ratio of the resulting
spectra..
We performed wavelength and flux calibration as well as geometrical
correction for two-dimensional distortion following the standard
reduction pipeline and applying the X2D task. This task corrected the
wavelength scale to the heliocentric frame too.
  
To measure the nuclear emission-line width we generally extracted
aperture spectra three ($0\farcs15$) and five pixels wide
($0\farcs25$) centered on the continuum peak, for the $0\farcs1$ and
$0\farcs2$-wide slit cases, respectively. When the spectra were
obtained with a 2-pixel binning read-out mode along the spatial
direction, we extracted aperture spectra three pixels wide
($0\farcs3$) for the $0\farcs2$-wide slit (Table~1).
The extracted spectra thus consist of the central emission convolved
with the STIS spatial point-spread function (PSF) and sampled over
 nearly square apertures of $0\farcs15\times0\farcs1$,
$0\farcs25\times0\farcs2$  or $0\farcs3\times0\farcs2$, roughly
corresponding to a circular aperture with  a radius of
$0\farcs07$, $0\farcs13$, and $0\farcs14$, respectively.
The wavelength range of our spectra is either 6482$-$7054 \AA\ or 6295
$-$ 6867 \AA, depending on whether the G750M grating was used at the
primary or secondary tilt.
The instrumental FWHM was 0.87 \AA\ ($\sigma_{\rm inst}= 17$ \kms) and
1.6 \AA\ ($\sigma_{\rm inst}= 32$ \kms) for the $0\farcs1$ and the
$0\farcs2$-wide slit, respectively.
The atlas of all the extracted spectra will be presented in a
forthcoming paper.

 To place our \mbh\ upper limits with the \ms\ relation, here we
consider only galaxies with velocity dispersion measurements in the
literature, which were available for 137 objects.
We also dropped a further five objects, since upon closer inspection
they revealed unrelaxed morphologies.
For a number of objects with a sharp central surface-brightness
profile, the two-dimensional rectification of the spectrum performed
during the data reduction produced anomalous undulations in the flux
level of continuum of the very central rows \citep[see][for
details]{KimQuijano2007}. This introduced also artificial fluctuations
in the emission-line flux profiles across the nucleus. As constraining
the concentration of the nebular emission is key to our modeling (see
\S\ref{sec:modeling}), this problem forced us to remove a further
eight galaxies from our sample.
  
\placefigone
  
\subsection{Measurement of the emission lines}  
\label{sec:measurement}  

In order to derive upper limits on \mbh\ following the method of
\citet[][see also \S\ref{sec:modeling}]{Sarzi2002} we need to
measure both the width of the central nebular emission and the radial
profile of the emission-line flux, so that we can gauge both the
depth of the potential well and the concentration of its gaseous
tracer.
To side-step the impact of broad and/or asymmetric emission
arising from regions much smaller than our resolution limit, we focus
on the width of the narrow component of the emission from forbidden
transitions and disregard the broad-line emission in our spectra. In
the wavelength range of our spectra that means measuring the central
width and flux profile of the \niipg\ lines since these are usually
brighter than the \siipg\ lines. The \nii\ doublet also traces the
nuclear kinematics better than \ha, given that this line could be
significantly affected by emission from circumnuclear star-forming
regions \citep[e.g.,][]{VerdoesKleijn2000, Coccato2006}.

To measure the central width and flux profile of the narrow component  
of the \nii\ lines we fit our spectra with multiple Gaussians to match  
both the broad and narrow components of all the observed lines, while  
describing the stellar continuum with a low-order polynomial. 
A flux ratio of 1:3 was assumed for the \nii\ doublet, as
dictated by atomic physics \citep[e.g.,][]{Osterbrock1989}, and in the
presence also of \sii\ emission, both the \nii\ and \sii\ doublets
were assumed to share a common line centroid and width.
In most cases only one additional Gaussian component was needed
in our fits, to describe the \ha\ emission from the broad-line region,
although in many objects also the forbidden \nii\ and \sii\ lines
required double-Gaussian profiles. This allowed us to describe also
lines with Voigt profiles, where tests on 18 galaxies showed that our
narrowest Gaussian component tend to be only slightly broader than the
thermal component in the Voigt profiles, generally by less than 20\%.
The extra Gaussian in the \nii\ and \sii\ lines was also used to
isolate the contribution of blue- or redshifted wings.
To help deblending the \ha+\nii\ region in some cases we followed
\citet{Ho1997b} and assigned to both the \nii\ lines and the narrow
\ha\ emission the line profile that was predetermined by fitting the
\sii\ lines.
The best-fitting Gaussian parameters were derived using a non-linear
least-squares minimization based on the robust Levenberg-Marquardt
method \citep[e.g.,][]{Press1996} implemented by \citet{More1980}. The
actual computation was done using the MPFIT algorithm\footnote{The
updated version of this code is available on
http://cow.physics.wisc.edu/$\sim$craigm/idl/idl.html} implemented by
C.~B. Markwardt under the IDL\footnote{Interactive Data Language is
distributed by Research System Inc.} environment.
In objects with conspicuous stellar absorption features that
cannot be accounted for by our minimisation routine we checked our
results against the line-width and flux measurements obtained with the
GANDALF software\footnote{The updated version of this code is
available at http://www.strw.leidenuniv.nl/sauron/software.html} of
\citet{Sarzi2006}, adopting either very young (300 Myr) or old (10
Gyr) stellar population templates. In most cases the measurements
agreed within the errors, except for IC~342 and NGC~7331 where
the \ha\ absorption line is particularly prominent. For these galaxies
we adopted the GANDALF values.
Finally, in defining our detection thresholds we compared the
amplitude ($A$) of the best-fitting line profile to the noise level
($N$) in the residuals of the continuum fit, adopting as detected only
those emission lines for which the $A/N$ ratio was larger than $3$.
Figure~\ref{fig:fitexample} shows a few sample spectra
illustrating the various fitting strategies explained above. A more
detailed description of our emission-line measurements for each of our
sample galaxy will be presented with our spectral atlas.

In 14 galaxies the nebular emission was too faint for it to be
detected given the quality of the corresponding spectra, and were
consequently dropped from our sample.
Three further galaxies had also to be discarded because their line
profile could not be well represented as a simple sum of Gaussian
components. Finally, two galaxies were rejected because the radial
profile of the flux of the \nii\ lines was strongly asymmetric and not
suitable for modelling.
Table~1 lists the final sample of galaxies analyzed in
this paper, which comprises of 105 galaxies which 28 ($27\%$) are
classified as ellipticals, 20 ($19\%$) are lenticulars, and 57
($54\%$) are spirals.
The central velocity dispersion of the ionized-gas component and the
size of the aperture we measured are also given in
Table~1.
Prior to modeling, the instrumental resolution corresponding to the
adopted apertures (17 \kms\ and 32 \kms\ for the $0\farcs1$ and
$0\farcs2$ slit widths, respectively) was subtracted in
quadrature from the observed line-width values to obtain the intrinsic
gas velocity dispersion.

Table~2 lists the 74 rejected galaxies.

\subsection{Modeling the central line width} 
\label{sec:modeling} 
 
Assuming that the width of the nuclear emission traces the depth of
the gravitational well, we can derive stringent upper bounds on the
mass of the SMBHs in our sample galaxies thanks to the exquisite
spatial resolution of HST. Although the stellar contribution to the
gravitational potential could affect such estimates, the fundamental
reason for which a lower limit on \mbh\ can not be set from such a
simple measurements is that the observed line-broadening may, in
principle, be entirely due to additional contributions such as
non-gravitational forces (e.g., gas pressure or magnetic
forces).

In this study we follow the procedure described in
\citet{Sarzi2002}, where a detailed description of the method can be
found.
In short, we assume that the observed line-broadening arises from the
motion of ionized-gas in a coplanar thin inner disk of unknown
inclination, where the gas moves in circular and Keplerian orbits
around the putative SMBHs.
For a given radial profile of the nebular emission, perfectly edge-on
disks lead to the broadest lines and therefore to a lower estimate of
\mbh. Conversely, the \mbh\ value needed to explain the observed line
width diverges to infinity as we approach perfectly face-on
configurations. Fortunately, such extreme orientations are
statistically rare.
Since randomly oriented disks have uniformly distributed $\cos{i}$, it
is possible to derive $1\sigma$ upper and lower limits on \mbh\ by
simply considering models with nearly edge-on ($i=81^\circ$,
$\cos{i}=0.16$) and face-on ($i=33^\circ$, $\cos{i}=0.84$)
orientations, respectively, comprising 68\% of the distribution of
\mbh\ values that can explain a given line width
\citep[e.g.,][]{Sarzi2002}.
 
In our models we could disregard the effect on the unknown position
angle of the disk since we extracted our spectra in nearly square
apertures, and thus assumed that the STIS slit was placed along the
disk major axis.
 
Clearly, for a given disk orientation the concentration of the gas
tracer impacts heavily on the \mbh\ value needed to explain a given
line width, to the point that no lower limit on \mbh\ can be set when
the gas profile is unresolved.
This is why the intrinsic emissivity distribution of the gaseous disk
has to be constrained from the data. As in \citet{Sarzi2002}, we
assumed an intrinsically Gaussian flux profile centered on the stellar
nucleus, which makes it easier to match the observed flux profile
while accounting for instrumental effects.
The choice of a Gaussian parametrization is also conservative, since
cuspier functions would have led us to estimate smaller \mbh. For
instance, adopting an exponential profile for the subsample of objects
studied also by \citet{Sarzi2002} leads on average to a $10\%$
decrease for the \mbh\ estimates.
 
In this work we disregarded the contribution of the stellar potential, 
which would lead to tighter upper limits on \mbh. 
In principle, it is possible to estimate the stellar mass contribution
by deprojecting the stellar surface brightness observed in the STIS
acquisition images while assuming spherical symmetry and a constant
mass-to-light ratio \citep{Sarzi2002}. In practice, however, this
would only be feasible for a limited number of objects in our sample,
given the limited quality of the acquisition images for most of our
sample galaxies, and the pervasive presence of dust absorption
features, in particular in spiral host galaxies.
Still, the impact of the stellar potential is unlikely to change 
dramatically our \mbh\ estimates, in particular for the upper limits 
derived for nearly face-on configurations. 
For their sample of nearby galaxies (at $8-17$ Mpc), \citet{Sarzi2002} 
found that including the stellar mass contribution reduced the median 
value of the \mbh\ upper limits by just $\sim12\%$. 
For the median distance our sample (21.4 Mpc) the stellar mass 
contribution to our $33^\circ$ upper-limits would be $\sim15\%$. 
Similar considerations would apply to the \mbh\ sensitivity limit
of our experiment. In the case of the \citet{Sarzi2002} sample this
value was found to be on average $3.9\times10^6$ \msun, which is well
below most of the \mbh\ limits derived here and comparable to the
smallest \mbh\ limits obtained for the closest objects in our sample.

To conclude, we note that the range spanned by the two \mbh\
values delivered by our Keplerian-disk model includes also \mbh\ that
would be estimated under radically different assumptions. For
instance, the gaseous disk model at $i=60^\circ$ is equivalent to that
of an isotropic gas sphere in hydrostatic equilibrium \citep[see][for
details]{Sarzi2002}.
The \mbh\ estimates we derived for $i=33^\circ$ and $81^\circ$ are
listed in Table~1 for the sample galaxies. Although
strictly speaking both values should be regarded as upper-limits, we
will refer only to the $33^\circ$\ estimates as \mbh\ upper limits,
hereafter.

\placefigtwo
 
\section{Results and discussion} 
\label{sec:results} 
 
We have determined the $1\sigma$ upper and lower confidence limits for
the \mbh\ for randomly orientated disks for 105 galaxies with
measurable spectra and stellar velocity dispersions
available in the literature.
For 19 galaxies of the sample, either \mbh\ measurements or upper
limits based on resolved kinematics were available
(Table~1).  Fig.~\ref{fig:ULcomparison} shows how such
measurements compare with our \mbh\ limits, once our values are
rescaled accordingly to the distances adopted in these studies. Our
\mbh\ upper-limits are consistent within 1$\sigma$ with such
estimates, except for NGC~3031 and NGC~4261.
Furthermore no systematic offset appears when our upper limits
are compared with similar upper bounds in the literature, rather than
definite measurements.
A particularly complex blend of narrow \ha+\nii\ and broad \ha\ lines
may have biased our \mbh\ estimates in NGC~3031.
 
To place our \mbh\ limits on the various versions of the \ms\
relation, we applied to the aperture correction of
\citet{Jorgensen1995} to the literature values of stellar velocity
dispersion in order to obtain the values \sigmac\ and \sigmae\
and that would have been measured within a circular aperture of radius
$r_{\rm e}/8$ and $r_{\rm e}$, respectively. The effective radii
$r_{\rm e}$ of the spheroidal components of our sample galaxies were
taken from various sources in the literature (Table~1)
except for few disk galaxies for which $r_{\rm e}$ was obtained from
our own photometric decomposition \citep[following][]{Mendez2008} of
the $K$-band images retrieved from the archive of the Two Micron All
Sky Survey \citep[][hereafter 2MASS]{2MASS}.

In Figs.~\ref{fig:ULmsigmaFF05} and \ref{fig:ULmsigmaLauer07} we
compare our \mbh\ upper limits to the \ms\ relation, as given by
both \citet{Ferrarese2005} and \citet{Lauer2007b}, initially to
establish the validity of our method over a wide range of velocity
dispersions.
Our upper bounds show a well defined trend with both \sigmac\ and
\sigmae, running closely above the \msc\ and \mse\ relations. In the
\msc\ plane a Spearman's rank coefficient of 0.9 suggests the presence
of a correlation at 9$\sigma$ confidence level whereas a Pearson
correlation coefficient of 0.8 supports a linear fit to the logaritmic
data, which returns a slope of $3.43\pm0.21$. At first glance such a
value would imply a shallower trend than found by
\citet{Ferrarese2005} and a slope closer to that of the
\citet{Lauer2007b} relation, but we need to keep in mind that the
derived slope could be significantly affected by just a few
outliers. In particular, for small value of $\sigma$ our \mbh\
upper-limits could be biased owning to a larger stellar contribution
to the gravitational potential in small and distant galaxies.
On the other hand, we found that our limits appear to parallel
particularly well both versions of the \ms\ relation for $90 \leq
\sigma_{\rm c}, \sigma_{\rm e} \leq 220$ \kms, whereas at lower and
higher $\sigma$ a substantial fraction of our \mbh\ limits lie either
considerably above or almost on top of the \ms\ relation,
respectively.
 
In the following sections we better quantify and interpret these first 
considerations. 
 
\subsection{Main trend in the sample}  
\label{subsec:coeff}  
  
In the \sigmac\ interval between $90$ and $220$ \kms\ our \mbh\ upper
limits appear to correlate particularly well with \sigmac, paralleling
the \msc\ relation.
In this \sigmac\ region a value of 0.8 for the Spearman's rank
correlation coefficient suggests the presence of a correlation at a
7-$\sigma$ confidence level, whereas a Pearson coefficient of 0.8
indicates that the logarithm values of our \mbh\ upper limits and
\sigmac\ are very likely to be linearly correlated.
A linear fit in the $\log{\sigma_{\rm c}}-\log{M_\bullet}$ plane
delivers a best-fitting slope of $4.52\pm0.41$ for our \mbh\ upper
limits, compared to the $4.86\pm0.43$ slope of the
\cite{Ferrarese2005} relation, with a scatter of 0.39 dex
(Figs.~\ref{fig:histograms}a).
In the $90 \leq \sigma_{\rm c} \leq 220$ \kms\ interval we have 66
\mbh\ upper limits, which have a median 2.7 times higher than the
expected \mbh\ value (Fig.~\ref{fig:histograms}c). These upper-limits
can range from falling short of the expected \mbh\ values by a factor
3.7 to exceeding them by a factor 17.3, although 68\% of them actually
do not top the expected \mbh\ values by more than a factor 4.1 and
fall immediately above the \msc\ relation.
For comparison, by fitting our upper limits in the \mse\ plane we
obtain a slope of $4.12\pm0.38$, very close to the value of
$4.13\pm0.32$ found by \citet{Lauer2007b}. In fact, the parallel trend
of our upper limits holds as far as $\sigma_{\rm e} \sim 300$ \kms,
with a Spearman's rank coefficient of 0.8, a Pearson linear
correlation coefficient of 0.8 and a linear slope of $3.84\pm0.28$.

\placefigthree
\placefigfour
\placefigfive

Fig.~\ref{fig:ULmsigmaFF05}a shows that such a trend holds independent
of galactic distance -- objects as far away as 60 Mpc appear to run
parallel to the \msc\ relation. In particular, the objects at and
below 20 Mpc are well distributed.
In fact, if in this range of \sigmac\ we perform separate linear
regression for the three different populations of upper limits with $D
< 30$ Mpc, $30< D <60$ Mpc and $60< D <100$ Mpc, we find slope values
that are consistent within the errors, namely of 4.05$\pm$0.51,
3.51$\pm$1.01, 4.52$\pm$1.14, respectively.
This finding shows that the observed nuclear line widths do not simply
trace an increasingly larger subtended stellar mass, as galaxies with
progressively larger stellar velocity dispersions are found
preferentially at larger distances. Instead, the fact that our upper
limits scale in the same way with $\sigma_c$ as precisely-measured
\mbh\ determinations indicates that the nuclear emission we measured
arises predominantly in regions of the gravitational potential that
are dominated by the influence of the central SMBHs.
  
This is not completely unexpected given that a number of HST
observations revealed that the narrow-line regions of active nuclei
appear to be quite concentrated with scales less than 50 pc, much more
so than the underlying stellar density profile \citep[see,
e.g.,][]{Ho2008}. Most recently, \citet{Walsh2008} have mapped the
behavior of the narrow-line region for galaxies observed with
multiple-slit STIS observations. They found that all galaxies of their
sample exhibit a centrally peaked surface-brightness profile, with the
majority of them further showing a marked gradient of the
emission-line widths within the sphere of influence of the central
SMBH.
The high degree of concentration of the gaseous tracer of the
gravitational potential is what allows us to closely trace the
presence of the SMBH even in objects where formally its sphere of
influence is not resolved. This is similar to the case of the
stellar-dynamical estimates of \mbh\ for M32, which have not
significantly changed when moving from ground- to space-based
observations \citep[e.g.,][and references therein]{Kormendy2004} due
to the exceptional concentration of its stellar light profile.
Actually, that fact that our upper limits run so closely to the \msc\
relations also suggests that non-gravitational forces do not generally
contribute much to the observed line widths (unless for some reason
their importance scales with \sigmac), although their role cannot be
firmly excluded on a single-case basis.
  
Fig.~\ref{fig:ULmsigmaFF05}b also shows that in the $\sigma_{\rm c} =
90 - 220$ \kms\ range the upper limits derived in galaxies of
different Hubble types lie neither closer nor further away from the
\msc\ relation, although only a relatively small number of elliptical
galaxies falls in this \sigmac\ interval. Similarly, even though only
$38\%$ of the spiral and lenticular galaxies in this \sigmac\ region
are unbarred we do not notice any systematic trend with the presence
of a bar, unlike what was found by \citet{Graham2008}.
  
Since our upper limits appear to trace quite closely the expected
values for \mbh, we can take advantage of the significant number of
galaxies in our sample to understand whether the objects that within
the present \sigmac\ range appear to show remarkably large or small
upper-bounds are in fact exceptional.
Assuming that our $1\sigma$ limits bracket symmetrically the expected
values of \mbh\ and that our upper bounds lie 2.7 times above the
\msc\ relation, with the aid of a Monte Carlo simulation we found that
$16\%$ of our \mbh\ upper limits should lie above the \msc\ relation
by more than 3 times its scatter \citep[adopting 0.34 dex
by][]{Ferrarese2005}, while $8\%$ of them should lie below it by more
than its scatter.
As Fig.\ref{fig:ULmsigmaFF05} shows, only four out of 66 objects in
the $\sigma_{\rm c} = 90 - 220$ \kms \ range fall that far above the
\msc\ relation, with an equal number falling below it by more than its
scatter. Both sets of objects correspond to $6\%$ of the galaxies in
the considered \sigmac\ range.

The previous considerations strongly argue against the presence of
exceedingly large \mbh\ (i.e., above the \msc\ relation by more than 3
times its scatter) in nearby galactic nuclei, and further suggests
that galaxies with considerably smaller \mbh-budgets (i.e., below the
\msc\ relation by more than its scatter) may be particularly rare.
The presence of undermassive SMBHs in field galaxies have been
suggested for instance by \cite{Vittorini2005}, who argued that in a
low galactic-density environment the \mbh\ growth may be hampered by
the lack of gaseous fuel. A population of undermassive SMBH was
discovered also by \citet{Volonteri2007} in her simulations of the
last stages of black-hole mergers, when the binary experiences a
recoil due to asymmetric emission of gravitational
radiation. According to Fig.~4 of \citet{Volonteri2007} up to $25\%$
of the galaxies with $90 \leq \sigma_{\rm c} \leq 220$ \kms\ could
contain undermassive SMBHs, with less than $10\%$ the expectected
\mbh.
Unfortunately, only a handful of objects in our sample are massive and
close enough (e.g., for $\sigma_{rm c} > 150$ \kms\ and $D< 20$ Mpc)
to allow us to probe such a low \mbh\ regime, where our simulations
indicate that we should expect only $\sim1$\% of our upper-limits.
Furthermore, although the wide range of Hubble types and of values for
\sigmac\ spanned by our sample galaxies suggest these are fairly
representative of the general properties of the nearby population, our
sample is almost certainly incomplete, in particular as we probe the
low-end of the luminosity function where most galaxies are found. Our
constraints should therefore be regarded with caution.
 
\subsection{The lower end of the \msc\ relation}  
\label{subsec:low}  
  
At small \sigmac\ ($< 90$ \kms) half of our upper limits
systematically exceed the expected \mbh\ values by 3 times the scatter
of the \msc\ relation. They are in average larger by more than a
factor 40 (Fig.~\ref{fig:histograms}d), consistent with previous works
on much more smaller samples \citep{Sarzi2002, Sarzi2004,
VerdoesKleijn2006}. They are hosted by NGC~3021, NGC~4245, NGC~5347,
NGC~5427, NGC~5879, and UGC~1395, which are late-type spirals with
different degrees of nuclear activity, as measured by \citet[][see
Table~1]{Ho1997}.
  
We have considered different possibilities related to the measurement
and modeling of the \niig\ emission line to explain the high values of
\mbh\ found in these 6 objects.
For instance, the presence of broad or asymmetric components in our
spectra could affect the width of the narrow component of the \nii\
lines that we measured and consequently the \mbh\ upper limits giving
larger masses. A similar bias would be introduced if the extent of the
flux profile were to be systematically overestimated.
Blue asymmetries are observed in the top outliers NGC~5347 and
NGC~5427, which are also part of the sample of active galactic nuclei
studied by \citet{Rice2006}, who investigated the resolved kinematics
of their narrow-line region with STIS spectra and first reported the
presence of blue wings in the \siipg\ lines. In our fits, however, the
contribution of such additional features was isolated using double
Gaussian profiles.
As regards the flux profile of the \nii\ doublet of the small-\sigmac\
outliers, these are not systematically shallower than the other
galaxies following the \msc\ relation in the same \sigmac\
range. Therefore, the \mbh\ upper limits that we have calculated are
not biased by either of these effects.
  
To explain the largest \mbh\ upper limits found at low \sigmac\ values
we also considered the impact of the presence of a nuclear star
cluster (NC), and in general that of a larger stellar contribution due
to a greater distance.
NCs are massive stellar clusters coincident with the galaxy
photocenter \citep{Cote2006} that are found in about $75\%$ of
late-type spiral galaxies \citep{Boker2002}. Their mean effective
radius is $\sim 3.5$ pc \citep{Boker2004}, small enough for them to be
completely enclosed within the central aperture of our spectra.
\citet{Ferrarese2006} found a different \msc\ relation for NCs, with
similar slope but a normalization that is larger by roughly an order
of magnitude than that the one found for SMBHs. The presence of NCs in
our low-\sigmac\ outliers could therefore explain why they show such
high central mass concentrations as indicated by their high \mbh\
values.
To assess the incidence of NCs in the sample galaxies with
$\sigma_{\rm c} < 90$ \kms\ we analyzed their surface-brightness
radial profile obtained with the IRAF task ELLIPSE on the STIS
acquisition images. For half of the low-\sigmac\ outliers we could
recognize the presence of a NC (NGC~3021, NGC~4245, and NGC~5879).
On the other hand, we could identify a NC only in one (NGC~4212) of
the six galaxies ($17\%$) which run close to the \msc\ relation
(IC~342, NGC~2685, NGC~2748, NGC~3982, NGC~4212, and NGC~5194).
The presence of a NC in the galaxies at the low-\sigmac\ end of our
sample is shown in Fig.~\ref{fig:ULmsigmaFF05}, and in the case of
NGC~3021 and and NGC~5879 it was already known
\citep[see][respectively]{Scarlata2004,Seth2008}.
If our limits trace indeed the dynamical signature of a NC in these
nuclei, better data and more detailed modelling
\citep[e.g.,][]{Barth2008} would be required to disentagle the
contribution of the NC and SMBH to the total mass budget.
As regards the distance of the low-\sigmac\ outliers, although we can
only rely on distances inferred from their recessional velocities it
is significant that half of them (NGC~5347, NGC~5427, and UGC~1395)
are found beyond 30 Mpc, whereas all the other low-\sigmac\ galaxies
are significantly closer, including those for which our \mbh\ upper
limits lie well within 3 times the scatter of \msc\ relation.

These findings suggest that part, if not all, of the exceedingly large
\mbh\ values we found at the low-\sigmac\ end of the \msc\ relation
could be ascribed to a more significant stellar contribution to the
gravitational potential. This is either because of the presence of a
nuclear stellar cluster (in NGC~3021, NGC~4245, and NGC~5879) or due
to a larger galactic distance (for NGC~5347, NGC~5427, and UGC~1395)
than otherwise required to trace the \msc\ relation at these \sigmac\
regimes.
Therefore, we presently do not need to invoke either non-gravitational
forces \citep[see][]{Sarzi2002} or a population of more massive SMBHs
\citep[see][]{Greene2006} to explain the observed flattening of the
\msc\ relation at low \mbh\ values.

\subsection{The upper end of the \msc\ relation}  
\label{subsec:high}  
  
At high \sigmac\ ($> 220$ \kms) our \mbh\ upper limits nicely bracket
the \msc\ relation (Fig.~\ref{fig:histograms}e) and most of them are
consistent with its scatter (Fig.~\ref{fig:ULmsigmaFF05}).
In fact, only four objects ($15\%$; NGC~2911, NGC~4552, NGC~4594, and
NGC~5077) fall above the \msc\ relation by more than its scatter, with
the same number of galaxies falling as far below the \msc\ relation
(NGC~3998, NGC~4278, NGC~6861, and UGC~1841).
These outliers do not stand out from the rest of the objects with
\sigmac$ > 220$ \kms\ for any obvious property such as morphology,
nuclear activity or distance.
This behavior is suggestive of an actual flattening of the high-mass
end of \msc\ relation, in particular considering that in the most
massive and radio-loud galaxies the ionized-gas velocity dispersion
can show a significant excess over a purely gravitational model
\citep[e.g.,][]{VerdoesKleijn2006}\footnote{In fact, for the two
radio-loud galaxies in our sample that were also studied by Verdoes
Kleijn et al. (NGC~383 and UGC~7115) the derived \mbh\ upper limits
lie above the \msc\ relation}.

The flattening at high-$\sigma$ values is less evident when our upper
limits are compared to the shallower \mse\ relation of
\citet{Lauer2007b}, but is nonetheless present upon closer inspection.
In particular, excluding objects with \sigmae$ < 90$ \kms\ where the
impact of the stellar potential on our \mbh\ estimates could be more
important, we found a systematic flattening in the main trend of our
upper limits as the high-\sigmae\ end of the \mse\ plane is
progressively populated. Specifically, whereas a linear fit to objects
with \sigmae$=90 - 220$ \kms\ yields a slope of $4.12\pm0.38$ (\S3.3),
extending this range to $280$ \kms, $340$ \kms\ and up to the maximum
\sigmae\ value in our sample of 386 \kms\ results in best-fitting
values of $3.86\pm0.29$, $3.78\pm0.27$ and $3.56\pm0.26$,
respectively.

This finding would be in agreement with the predictions of
semi-analytic models for the coevolution of SMBHs and galaxies at the
highest end of the mass spectrum, whereby galaxies and SMBHs grow
mainly via gas-poor, dry mergers \citep{Schawinski2006}.
Yet, the behavior of \msc\ relation in this regime is still under
debate. In particular, the limited number of galaxies with reliable
\mbh\ measurement in the range \mbh $ > 10^9$ \msun\ are actually
consistent with a steepening of the \msc\ relation
\citep[e.g.,][]{Wyithe2006,DallaBonta2008}.
Furthermore, the cutoff at $\sigma_{\rm c} \sim 400$ \kms\ of the
local velocity dispersion function \citep{Seth2008} implies either
that SMBHs with \mbh$>3 \times 10^9$ \msun\ are extremely rare or that
if they exists their host galaxies should lay considerably above the
present \msc\ relation.
In fact, at these regimes \citet{Lauer2007} argue that the stellar
luminosity $L$ is better suited than \sigmac\ to trace \mbh.  The
\msc\ relation should steepen at its high-\sigmac\ end if Lauer et
al. arguments are correct, since the observed \sigmac\ saturates for
the most massive of ellipticals while considering increasingly large
values of $L$.

Although our results suggest a flattening of the \msc\ relation,
we need to keep in mind that systematic effects related to the
measurement of the bulge properties may be significant at the
high-\sigmac\ end of the \msc\ plane. In particular, the
aperture correction for the stellar velocity dispersion may be both
more important and more uncertain for the most massive of ellipticals
than for smaller elliptical and lenticular galaxies.
Indeed, giant ellipticals tend to have shallower central surface
brightness profiles than their less massive counterparts, which makes
the aperture correction more sensitive to the quality and spatial
coverage of the stellar kinematics and to uncertainties on the value
of the galaxy effective radius, $r_{\rm e}$. Incidentally,
measurements or $r_{\rm e}$ are also generally less accurate for giant
ellipticals, due the presence of extended stellar halos.
Ideally, rather than \sigmac\ one would like to have a quantity that
is more closely connected to the stellar mass, such as the total
$K$-band luminosity, which is also known to relate to \mbh\
\citep{Marconi2003a}, or a {\it direct} measurement of \sigmae.
Obtaining the $K$-band luminosity of our sample galaxies would require
much deeper images than the available 2MASS data, whereas properly
measuring \sigmae\ requires integral-field observations, such as those
derived in the case of the SAURON survey \citep{Emsellem2007}.

\subsection{Summary}  
\label{sec:concl}  
  
Owing to the exquisite spatial resolution of HST and to the
concentrated character of the ionized-gas emission in low-luminosity
AGNs, we have been able to set tight upper limits on \mbh\ for a
sample of 105 nearby galaxies ($D<100\,\rm Mpc$) using STIS/G750M
spectra. This sample spans a wide range of Hubble types (with 54\% of
spirals) and includes objects with published values for their central
stellar velocity dispersion \sigmac.
Our main findings are: 
   
\begin{itemize}  
\item Independent of the galaxy distance, our \mbh\ upper limits run 
      parallel and above the \msc\ relation, in particular for values 
      of \sigmac\ between 90 and 220 \kms. The median of the 66 \mbh\ 
      upper limits in this \sigmac\ regime exceeds the expected \mbh\ 
      value by a factor 2.7, with 68\% of our upper limits falling 
      immediately above the \msc\ relation and without exceeding the 
      expected \mbh\ values by more than a factor 4.1. 
 
\item That our nebular line-width measurements trace rather well the
      nuclear gravitational potential, makes large samples of \mbh\
      upper-limit measurements useful in constraining the frequency of
      objects with exceedingly low or high values of \mbh\ and in
      probing the black-hole mass budget across the entire Hubble
      sequence.
       
\item No systematic trends or offsets are observed in this \sigmac\ 
      range as a function of the galaxy Hubble type, or with respect to 
      the presence of a bar. 
      Furthermore, no evidence was found to suggest that the largest
      or smallest \mbh\ upper limit in the \sigmac\ range between 90
      and 220 \kms\ are actually bracketing exceptionally high or low
      values of \mbh. Thus, galaxies with exceedingly high \mbh\
      budgets must be very rare.
  
\item For \sigmac\ values below 90 \kms\ half of our \mbh\ upper  
      limits systematically exceed the expected \mbh\ values by more  
      than a factor 40, consistent with previous work on much smaller  
      samples. 
     
      The line-width measurements for such low-\sigmac\ outliers are 
      most likely affected by the stellar contribution to the 
      gravitational potential, either due to the presence of a nuclear 
      stellar cluster or because of a greater distance compared to the 
      other galaxies at the low-\sigmac\ end of the \msc\ relation, for 
      which our \mbh\ upper limits are closer to the expected \mbh\ 
      values. 
  
\item At the opposite \sigmac\ end of the \msc\ relation, for values of 
      \sigmac\ above 220 \kms, our \mbh\ upper bounds appear to lie 
      much closer the expected \mbh\ in the most massive elliptical 
      galaxies, even falling below the \msc\ relation. 
      This flattening is less evident when our upper limits are
      compared with the shallower \mse\ relation by
      \citet{Lauer2007b}, but is nonetheless present upon closer
      inspection.  In particular, excluding objects with \sigmae$ <
      90$ \kms , we found a systematic flattening in the main trend of
      our upper limits as the high-\sigmae\ end of the \mse\ plane is
      progressively populated.
      
      Although such a flattening of the \msc\ relations at its 
      high-\sigmac\ end would appear consistent with models for the 
      coevolution of supermassive black holes and galaxies driven by 
      dry mergers, we caution that better and more consistent 
      measurements for either the $K$-band luminosity or the 
      integrated value of the stellar velocity dispersion \sigmae\ 
      within the bulge effective radius $r_{\rm e}$ (both better 
      tracers of the bulge mass than \sigmac) are needed before 
      systematic effects can be ruled out.\\

\end{itemize}

\bigskip   
\noindent   
{\bf Acknowledgments.}    
We acknowledge the anonymous referee for his/her many comments that
improved our manuscript. We are indebted with James Binney, St\'ephane
Courteau, Sadegh Khochfar, Lorenzo Morelli, Kevin Schawinski, and
Massimo Stiavelli for many useful discussions and suggestions.  We are
also indebted with Jonelle Walsh and her collaborators for sharing
with us the results of their work prior to publication.  We thank
Jairo M\'endez Abreu for the GASP2D package which we used for
measuring the photometric parameters of some of the sample galaxies.
We acknowledge the grant CPDA068415/06 by Padua University, which
provided support for this research. AB is grateful to the University
of Hertfordshire for its hospitality while this paper was in progress.
This research has made use of the Lyon-Meudon Extragalactic Database
(LEDA), NASA/IPAC Extragalactic Database (NED), and the Two
Micron All Sky Survey (2MASS).

\newpage 
 
\placefigsix
\placefigseven

\clearpage

\begin{center} 
\begin{scriptsize} 
\begin{minipage}{16cm} 

{\sc Notes.} ---  
Col.(1): Galaxy name.  
Col.(2): Morphological type from RC3. 
Col.(3): Nuclear spectral class from \citet{Ho1997}, where H = H~II 
nucleus, L = LINER, S = Seyfert , T = transiton object (LINER/HII), 1 
= type 1, 2 = type 2, and a fractional number between 1 and 2 denotes 
various intermediate types; uncertain and highly uncertain 
classifications are followed by a single and double colon, 
respectively. The nuclear spectral class of galaxies marked with * is 
from NASA/IPAC Extragalactic Database (NED). 
Col.(4): Distance. 
Col.(5): Reference for col. 4. All the distances were taken from 
literature (see attached list), except those we obtained from $V_{\rm 
3K}$, the weighted mean recessional velocity corrected to the reference 
frame of the microwave background radiation given in RC3. These were 
derived as $V_{\rm 3K}/H_0$ with $H_0 = 75$ km s$^{-1}$ Mpc$^{-1}$. 
Col.(6): Absolute corrected $B$ magnitude derived from $B^0_T$ (RC3) 
with the adopted distance. 
Col.(7): Effective radius of the spheroidal component.  

Col.(8): Band in wich the effective radius were measured.  
Col.(9): Reference for col.(7). All the effective radii were taken 
from literature (see attached list), except for those we measured by a 
photometric decomposition of the $K$-band images available in the 
2MASS science archive \citet{2MASS}. 
Col.(10): Central velocity dispersion of the stellar component within 
$r_e/8$. 
Col.(11): Reference for the measured stellar velocity dispersion and 
corresponding size of the central aperture from which we calculated 
the value given in col.(10) by following \citet{Jorgensen1995}. We did 
not apply any aperture correction to the measured stellar velocity 
dispersions of NGC~2748, NGC~3982, and UGC~1841, because no 
information about the size of the aperture was available. 
Col.(12): HST proposal number under which was obtained the STIS/G750M 
spectrum from which we measured the central velocity dispersion of the 
ionized gas. 
Col.(13): Size of the central aperture where we measured the velocity 
dispersion of the ionized gas. 
Col.(14): Physical size of the central aperture where we measured the velocity 
dispersion of the ionized gas.
Col.(15): Central velocity dispersion of the ionized-gas
  component within the aperture in Col. (13).  This is the intrinsic
  velocity dispersion obtained from the observed one by subtracting
  the instrumental velocity dispersion. 
Col.(16): \mbh\ upper limit for a Keplerian disk model assuming 
$i=33^\circ$. 
Col.(17): \mbh\ upper limit for $i=81^\circ$. 
Col.(18): Mass (and confidence interval) of the SMBH derived from
modeling based on the resolved kinematics. The \mbh\ of NGC~3227 and
NGC~4258 were obtained by studying the dynamics of stars and water
masers, respectively. The ionized-gas dynamics was used for all the
remaining galaxies.
Col.(19): Reference for col.(18). 
 
{\sc References.} --- 
(\citetalias{Tully1988})		   \citet{Tully1988};		  
(\citetalias{RC3})		   \citet{RC3};		  
(\citetalias{Tonry2001})		   \citet{Tonry2001};		  
(\citetalias{Freedman2001})	   \citet{Freedman2001};  
(\citetalias{2MASS})      	   \citet{2MASS};  
(\citetalias{Kassin2006})		   \citet{Kassin2006};		  
(\citetalias{deSouza2004})	   \citet{deSouza2004};	  
(\citetalias{Laurikainen2004})	   \citet{Laurikainen2004};	  
(\citetalias{Xanthopoulos1996})      \citet{Xanthopoulos1996}; 
(\citetalias{Baggett1998})	   \citet{Baggett1998}; 
(\citetalias{Andredakis1994})        \citet{Andredakis1994};      
(\citetalias{Marconi2003a})	   \citet{Marconi2003a};  
(\citetalias{FalconBarroso2002})	   \citet{FalconBarroso2002};	  
(\citetalias{Laurikainen2005})	   \citet{Laurikainen2005};	  
(\citetalias{Fisher1995})		   \citet{Fisher1995};	  
(\citetalias{Scarlata2004})	   \citet{Scarlata2004};  
(\citetalias{Terlevich1990})	   \citet{Terlevich1990};	  
(\citetalias{GarciaRissmann2005})	   \citet{GarciaRissmann2005};	  
(\citetalias{Wegner2003})		   \citet{Wegner2003};		  
(\citetalias{Batcheldor2005})	   \citet{Batcheldor2005};	  
(\citetalias{Davies1987})		   \citet{Davies1987};		  
(\citetalias{Simien1997a})	   \citet{Simien1997a};	  
(\citetalias{Bernardi2002})	   \citet{Bernardi2002};  
(\citetalias{Schechter1983})	   \citet{Schechter1983};	  
(\citetalias{Bonfanti1995}) 	   \citet{Bonfanti1995};	  
(\citetalias{Nelson1995})		   \citet{Nelson1995};		  
(\citetalias{Barth2002})		   \citet{Barth2002};		  
(\citetalias{Corsini1999})	   \citet{Corsini1999};	  
(\citetalias{Simien1997b})	   \citet{Simien1997b};	  
(\citetalias{Simien1997c})	   \citet{Simien1997c};	  
(\citetalias{Heraudeau1998})	   \citet{Heraudeau1998};	  
(\citetalias{diNella1995})	   \citet{diNella1995};	  
(\citetalias{Heraudeau1999})	   \citet{Heraudeau1999};	  
(\citetalias{VegaBeltran2001})	   \citet{VegaBeltran2001};	  
(\citetalias{Carollo1993})	   \citet{Carollo1993};	  
(\citetalias{Simien1998})		   \citet{Simien1998};		  
(\citetalias{Whitmore1979})	   \citet{Whitmore1979};  
(\citetalias{Smith2000})		   \citet{Smith2000};		  
(\citetalias{Sarzi2002})		   \citet{Sarzi2002};		  
(\citetalias{Fisher1997})		   \citet{Fisher1997};		  
(\citetalias{Simien2002})		   \citet{Simien2002};		  
(\citetalias{FalconBarroso2006})	   \citet{FalconBarroso2006};	  
(\citetalias{VerdoesKleijn2002})	   \citet{VerdoesKleijn2002};	  
(\citetalias{Whitmore1981})	   \citet{Whitmore1981};  
(\citetalias{Jarvis1988})		   \citet{Jarvis1988};		  
(\citetalias{Proctor2002})	   \citet{Proctor2002};	  
(\citetalias{Bettoni1997})	   \citet{Bettoni1997};	  
(\citetalias{Bender1994})		   \citet{Bender1994};		  
(\citetalias{NoelStorr2007})	   \citet{NoelStorr2007};	  
(\citetalias{Dumas2007})		   \citet{Dumas2007};		  
(\citetalias{Gu2006})		   \citet{Gu2006};		  
(\citetalias{Wagner1988})		   \citet{Wagner1988};		  
(\citetalias{vandenBosch1995})	   \citet{vandenBosch1995};	  
(\citetalias{Balcells1995})	   \citet{Balcells1995};   
(\citetalias{Tonry1981})		   \citet{Tonry1981};		  
(\citetalias{Atkinson2005})	   \citet{Atkinson2005}; 
(\citetalias{Sarzi2001})		   \citet{Sarzi2001};		  
(\citetalias{Devereux2003})	   \citet{Devereux2003};  
(\citetalias{Davies2006})		   \citet{Davies2006};		  
(\citetalias{Barth2001})		   \citet{Barth2001};		  
(\citetalias{Pastorini2007})	   \citet{Pastorini2007};	  
(\citetalias{deFrancesco2006})	   \citet{deFrancesco2006};	  
(\citetalias{Marconi2003b})	   \citet{Marconi2003b};  
(\citetalias{Miyoshi1995})	   \citet{Miyoshi1995};	  
(\citetalias{Ferrarese1996})	   \citet{Ferrarese1996};	  
(\citetalias{Coccato2006})	   \citet{Coccato2006};	  
(\citetalias{Macchetto1997})	   \citet{Macchetto1997};	  
(\citetalias{Kormendy1988})	   \citet{Kormendy1988};  
(\citetalias{deFrancesco2008})	   \citet{deFrancesco2008};	  
(\citetalias{vanderMarel1998})	   \citet{vanderMarel1998}; 
(\citetalias{Bower1998})	   \citet{Bower1998};
(\citetalias{Koprolin2000})	   \citet{Koprolin2000}.

\end{minipage}  
\end{scriptsize} 
\end{center}

\newpage

\begin{deluxetable}{l l c c}
\tablecolumns{4}
\tablenum{2}
\tablewidth{0pc}
\tablecaption{Properties of the galaxies rejected from the main sample\label{tab:rej}}
\tablehead{
\colhead{Galaxy} &
\colhead{Morp. T.} &
\colhead{Prop.} & 
\colhead{Rej.} \\
\multicolumn{1}{c}{} &
\multicolumn{1}{c}{(RC3)} &
\multicolumn{1}{c}{} &
\multicolumn{1}{c}{} \\
}
\startdata
IC~5096  &    Sbc sp        & 9046 & ns  \\ 
NGC~0134 &    SABbc(s)      & 8228 & ns  \\
NGC~0157 &    SABbc(rs)     & 8228 & ns  \\
NGC~0255 &    SABbc(rs)     & 8228 & ns  \\ 
NGC~0278 &    SABb(rs)      & 7361 & s   \\
NGC~0520 &    pec           & 8669 & i   \\ 
NGC~1097 &    SBb(s)        & 9782 & ds  \\
NGC~1255 &    SABbc(rs)     & 8228 & n   \\ 
NGC~1300 &    SBbc(rs)      & 8228 & p   \\ 
NGC~1832 &    SBbc(r)       & 8228 & ns  \\
NGC~2623 &    pec           & 8669 & i   \\
NGC~2654 &    SBab sp:      & 9046 & s   \\
NGC~2892 &    E$^+$ pec:    & 8236 & s   \\
NGC~2976 &    SAc pec       & 8591 & s   \\ 
NGC~3003 &    Sbc?          & 8228 & s   \\
NGC~3049 &    SBab(rs)      & 7513 & s   \\ 
NGC~3067 &    SABab(s)      & 8596 & f   \\ 
NGC~3162 &    SABbc(rs)     & 8228 & r   \\ 
NGC~3254 &    SAbc(s)       & 8228 & n   \\
NGC~3256 &    pec           & 8669 & i   \\ 
NGC~3259 &    SABbc(rs)     & 8228 & s   \\
NGC~3379 &    E1            & 8589 & f   \\ 
NGC~3403 &    SAbc:         & 8228 & ns  \\
NGC~3489 &    SAB0$^+$(rs)  & 7361 & p   \\ 
NGC~3516 &    (R)SB0(s)     & 8055 & d \\ 
NGC~3521 &    SABbc(rs)     & 8228 & ns  \\ 
NGC~3684 &    SAbc(rs)      & 8228 & ns  \\
NGC~3686 &    SBbc(s)       & 8228 & s   \\
NGC~3705 &    SABab(r)      & 8607 & n   \\ 
NGC~3756 &    SABbc(rs)     & 8228 & ns  \\
NGC~3887 &    SBbc(r)       & 8228 & s   \\ 
NGC~3917 &    SAcd:         & 8607 & ns  \\ 
NGC~3921 &(R$'$)SA0/a(rs) pec& 8669& i   \\ 
NGC~3949 &    SAbc(s)       & 8228 & ns  \\
NGC~3972 &    SAbc(s)       & 8228 & ns  \\
NGC~4030 &    SAbc(s)       & 8228 & r   \\
NGC~4038 &    SBm(s) pec    & 8669 & is  \\                
NGC~4039 &    SBm(s) pec    & 8669 & is  \\                
NGC~4051 &    SABbc(rs)     & 8228 & r   \\                
NGC~4100 &    (R$'$)SAbc(rs)& 8228 & s   \\
NGC~4138 &    SA0$^+$(rs)   & 1039 & n   \\ 
NGC~4303 &    SABbc(rs)     & 8228 & p   \\
NGC~4343 &    SAb(rs)       & 9068 & s   \\ 
NGC~4380 &    SAb(rs):?     & 7361 & n   \\
NGC~4389 &    SBbc(rs)      & 8228 & ns  \\
NGC~4414 &    SAc(rc)?      & 8607 & n   \\ 
NGC~4420 &    SBbc(r):      & 8228 & s   \\
NGC~4527 &    SABbc(s)      & 8607/8228 & fn \\
NGC~4536 &    SABbc(rs)     & 8228 & r   \\
NGC~4569 &    SABab(rs)     & 8607 & r   \\ 
NGC~4676A&    S0 pec ? (Irr)& 8669 & s   \\        
NGC~4696 &    E$^+$1 pec    & 8690 & n   \\ 
NGC~5054 &    SAbc(s)       & 8228 & s   \\
NGC~5055 &    SAbc(rs)      & 7361/8228 & n  \\ 
NGC~5135 &    SBab(s)       & 9143 & r   \\
NGC~5141 &    S0            & 8236 & s   \\
NGC~5247 &    SAbc(s)       & 8228 & ns  \\
NGC~5364 &    SAbc(rs) pec  & 8228 & n   \\
NGC~5398 &(R$'$)SBdm(s): pec& 7513 & s   \\ 
NGC~5577 &    SAbc(rs)      & 8228 & ns  \\
NGC~5635 &    S pec         & 7354 & s   \\
NGC~5713 &    SABbc(rs)     & 8228 & s   \\
NGC~5746 &    SABb(rs) sp:  & 9046 & n   \\
NGC~5905 &    SBb(r)        & 9177 & d   \\ 
NGC~5921 &    SBbc(r)       & 8228 & s   \\
NGC~6384 &    SABbc(r)      & 8228 & n   \\
NGC~6503 &    SAcd(s)       & 8607 & n   \\ 
NGC~6621 &    Sb: pec       & 8669 & s   \\
NGC~7252 &    (R)SA0$^0$:   & 8669 & i   \\
NGC~7314 &    SABbc(rs)     & 8228 & pr  \\
NGC~7592 &    S0$^+$ pec:   & 8669 & s   \\  
UGC~10814&    Scd:          & 9782 & ds  \\
\enddata
\end{deluxetable} 

\clearpage

\begin{center}
\begin{scriptsize}
\begin{minipage}{16cm} 

{\sc Notes.} ---  
Col.(1):  Galaxy name.  
Col.(2): Morphological type from RC3. 
Col.(3): HST proposal number under which was obtained the STIS/G750M 
spectrum. 
Col.(4): Reason of rejection, where 
d = problem in deblending the emission lines,
i = interacting galaxy,
f = irregular or strongly asymmetric radial profile of the flux of the \nii\ emission linea 
n = faint or absent emission lines,
p = double-peaked emission lines,
r = unsuccessful two-dimensional rectification of the spectrum,
s = no stellar velocity dispersion available in literature.

\end{minipage}  
\end{scriptsize}
\end{center}

\end{document}